\documentclass[fleqn,usenatbib,nofootinbib]{mnras}

\usepackage{newtxtext,newtxmath}
\usepackage[T1]{fontenc}

\usepackage{graphicx}	% Including figure files
\usepackage{amsmath}	% Advanced maths commands
\usepackage{longtable}
\usepackage{booktabs} 
\usepackage{multirow}
\usepackage{todonotes} 
\usepackage{adjustbox}
\usepackage{booktabs} 
\usepackage{longtable}
\usepackage{lscape}
\usepackage{array}
\usepackage{bm}
\usepackage{cprotect}
\makeatletter
\renewcommand{\@journal}{}
\newcommand{\verba}[1]{\texttt{\string#1}}
\makeatother

\title[]{Probabilistic cosmological inference on HI tomographic data}

\author[Andrianomena et al.]{Sambatra Andrianomena$^{1,2}$\thanks{E-mail: hagatiana.andrianomena@gmail.com}
\\
$^{1}$SARAO, Liesbeek House, River Park Liesbeek Parkway, Settlers Way, Mowbray, Cape Town, 7705 \\
$^{2}$Department of Physics \& Astronomy, University of the Western Cape, Bellville, Cape Town 7535,
South Africa\\
}

\pubyear{2015}

\begin{document}
\label{firstpage}
\pagerange{\pageref{firstpage}--\pageref{lastpage}}
\pagestyle{plain}
\maketitle

\begin{abstract}
We explore the possibility of retrieving cosmological information along with its inherent uncertainty from 21-cm tomographic data at intermediate redshift. The first step in our approach consists of training an encoder, composed of several three dimensional convolutional layers, to cast the neutral hydrogen 3D data into a lower dimension latent space. Once pre-trained, the featurizer is able to generate 3D grid representations which, in turn, will be mapped onto cosmology ($\Omega_{\rm m}$, $\sigma_{8}$) via likelihood-free inference. For the latter, which is framed as a density estimation problem, we consider a Bayesian approximation method which exploits the capacity of Masked Autoregressive Flow to estimate the posterior. It is found that the representations learned by the deep encoder are separable in latent space. Results show that the neural density estimator, trained on the latent codes, is able to constrain cosmology with a precision of $R^2 \ge 0.91$ on all parameters and that most of the ground truth of the instances in the test set fall within $1\sigma$ uncertainty. It is established that the posterior uncertainty from the density estimator is reasonably calibrated. We also investigate the robustness of the feature extractor by using it to compress out-of-distribution dataset, that is either from a different simulation or from the same simulation but at different redshift. We find that, while trained on the latent codes corresponding to different types of out-of-distribution dataset, the probabilistic model is still reasonably capable of constraining cosmology, with $R^2 \ge 0.80$ in general. This highlights both the predictive power of the density estimator considered in this work and the meaningfulness of the latent codes retrieved by the encoder. We believe that the approach prescribed in this proof of concept will be of great use when analyzing 21-cm  data from various surveys in the near future.
% Various cases where the posterior estimator is trained on representations are extracted by the encoder from dataset that is out-to-distribution to it are also investigated.
% allows the estimation of the posterior uncertainty reflecting the ignorance of the probabilistic model.  
\end{abstract}


%%%%%%%%%%%%%%%%%%%%%%%%%%%%%%%%%%%%%%%%%%%%%%%%%%

%%%%%%%%%%%%%%%%% BODY OF PAPER %%%%%%%%%%%%%%%%%%

\section{Introduction}

\cite{bull2015late} developed a framework, based on Fisher forecast formalism, which highlighted the potentials of HI intensity mapping \citep[HI IM;][]{battye2004neutral} surveys at low and intermediate redshifts to constrain cosmological parameters. Their analyses investigated the measurements of observables under different configurations from several HI experiments, such as MeerKAT \citep{jonas2009meerkat}, BINGO \citep{battye2013h}, CHIME \citep{bandura2014canadian}, SKA \citep{mellema2013reionization} and FAST \citep{smoot201721}. Similar study  was also carried out in \cite{pourtsidou2017h} where the cross-correlation between HI IM and optical galaxy surveys was explored. Complementing the constraints obtained from galaxy surveys such as the dark energy spectroscopic instrument \citep[DESI;][]{martini2018overview} and Sloan Digital Sky Survey \citep[SDSS;][]{dawson2016sdss}; and cosmic microwave background (CMB) such as Planck \citep{ade2014planck, aghanim2020planck}, we are essentially entering the era of precision cosmology by exploiting the great potential of HI IM, a powerful cosmological probe.

% According to the type of datasets obtained from HI surveys, there mainly exist three modalities.
There exist various modalities that can be considered in terms of extracting cosmological information. \cite{greig201521cmmc}, for instance, prescribed a pipeline which utilized a Bayesian framework in order to analyze the 21-cm power spectrum during the Epoch of Reionization (EoR). Using HI IM as a cosmological probe within the redshift range of $0.25 - 2.75$, \cite{ansari201221} studied the dependency of the extracted Large Scale Structure power spectrum $P(k)$ on different radio interferometer setups. They also constrained dark energy from a 21-cm Baryon Acoustic Oscillation (BAO) survey. With the ever increasing use of machine learning (ML) as a tool to analyze cosmological datasets, \cite{novaes2024cosmological} opted for a neural network model to map angular power spectrum unto cosmological parameters. Employing $P(k)$ information, which describes the clustering of matter at different scales, in cosmological analyses is a sensible approach. Nevertheless, the latter is subject to information loss which is potentially more severe on non-linear scales. In addition to that, 21-cm signal is known to be non-gaussian \citep[see for example][]{majumdar2018quantifying, greig2023detecting}. Utilizing higher order statistics, such as bispectrum, is one way to account for non-linearities \citep[see for instance][]{bharadwaj2005probing}, but it is not a trivial task. Instead of making use of lossy summary statistics to retrieve the information, another data modality, which is processed by a ML model for field level inference, is HI 2D map which is characterized by its resolution, in other words its number of pixels. \cite{gillet2019deep} demonstrated how a convolutional neural network (CNN) was able to recover parameters of the first galaxies during EoR. \cite{hassan2020constraining} investigated the impact of training their CNNs using noisy HI maps on parameter recovery, as a way to mimic real world scenarios where the data are expected to be contaminated by systematics. \cite{andrianomena2023predictive} used Monte Carlo Dropout \citep{gal2016dropout} (which consists of activating dropout layers in deep learning architecture during inference time) on their deep network to estimate the total uncertainty associated with each parameter prediction. They considered HI maps in their training datasets, but also a combination of HI with other fields (or also maps) like gas density and amplitude of the magnetic field. Another promising approach involves learning the feature representation of a 2D map, which can then be used for other downstream tasks such as regression. \cite{andrianomena2023latent}, as an example, employed a very deep variational autoencoder \citep{child2020very} to learn the latent codes of three channel images comprising gas density, neutral hydrogen and amplitude of magnetic field maps. They showed that the meaningful representations could be used to recover cosmological parameters. 
% . The latter is similar to a summary statistic but is composed of the salient features  

In the near future, the Hydrogen Intensity and Real-time Analysis eXperiment \citep[HIRAX;][]{crichton2022hydrogen} will produce a tomographic measurement of cosmological HI emission within the redshift range of $0.755 < z < 2.55$,  covering 15,000 deg$^2$ in the southern part of the sky. Similarly, in the northern sky, CHIME \citep{amiri2022overview} will provide large scale structure (LSS) HI distribution within the range $z = 0.8 - 2.5$. One of the science cases of SKA \citep{wagg2014ska1, mellema2015hi} is to also map the 3D HI distribution during EoR and Cosmic Dawn. In preparation for the upcoming tomographic HI data which encode a wealth of information, previous studies started to explore deep learning techniques that can potentially handle those 3D data, whether at low/intermediate redshift or high redshift during EoR.
% Detection of cosmological HI emission using CHIME \cite{amiri2023detection}
For instance, \cite{zhao2022simulation} utilized a 3D CNN combined with a likelihood free-inference method, PyDELFI \citep{alsing2019fast}, to harness the 3D information from HI lightcone data from EoR. They showed that their method improved on previous analysis based on HI 2D maps. \cite{zhao2024simulation} demonstrated that using solid harmonic wavelet scattering transform \citep{eickenberg2017solid, eickenberg2018solid} as a data compressor resulted in tighter credible regions of parameters than those obtained in \cite{zhao2022simulation}. \cite{binnie2025likelihood} also carried out a model selection between different reionization scenarios using a Bayesian framework. They first compressed the HI lightcone data by using a 3D CNN model and trained a modified version of PyDELFI on the summarized data to estimate the evidence. 
% \cite{mao2008accurately} investigated the accuracy that can be achieved leveraging 3D HI data. 
% By considering an HI intensity mapping survey at low redshifts, BINGO \citep{battye2013h}. 
% \cite{andrianomena2025towards} (domain adaptation). 
% \cite{jo2025towards} (robustness using HI data). 
% \cite{shao2023robust} inference from 3D galaxy catalogue, robust field inference level.

This far, to the best of our knowledge, recovering cosmological parameters from 21-cm tomographic data-cube at low/intermediate redshift using deep learning method hasn't been addressed. Based on the detection of neutral hydrogen by CHIME \citep{amiri2023detection} within $z = 0.78 - 1.43$, and the analysis conducted by \cite{ansari201221} at $z \sim 1$, our investigation is focused on $z = 1$. In this study, building on previous works \citep{zhao2022simulation, zhao2024simulation, binnie2025likelihood}, we investigate an approach that can help retrieve meaningful representations of otherwise high dimensional inputs, using much deeper architecture than those previously considered. We then build a mapping between cosmological parameters and the informative featurized data using a Bayesian approximation which doesn't assume a tractable form of likelihood. The density estimator used in this work provides epistemic uncertainty, which reflects the regression model ignorance.  
% A Bayesian approximation, which doesn't assume a tractable form of likelihood and is trained on the latent codes, is used to infer the cosmological parameters along with their corresponding posterior uncertainties.
This paper is organized as follows: we present the datasets in our analyses in \S\ref{sec:data}, and elaborate on the techniques employed in this work in \S\ref{sec:method}, we discuss the results in \S\ref{sec:results} and finally conclude in \S\ref{sec:conclusion}.

\section{Data}\label{sec:data}
As the objective of our study is a robust retrieval of cosmological information from HI tomographic data, we make use of the 3D grids of the multifield dataset \citep{villaescusa2022camels} of the CAMELS Project \citep{villaescusa2021camels,pacodatarelease}. Using data from the Green Bank Telescope, \cite{switzer2013determination} made a measurement of the HI fluctuations $\Omega_{\rm HI}b_{\rm HI}$, where $\Omega_{\rm HI}$ and $b_{\rm HI}$ are the HI density parameter and HI bias respectively,  at $z\sim 1 $. \cite{ansari201221} also predicted that the constraints on Baryon Acoustic Oscillations (BAO)  at $z\sim 1$ using radio interferometer data would be competitive with those obtained from optical surveys. Similarly, in our analyses, we focus on 3D grids at $z = 1$. However, to demonstrate the robustness of our parameter inference, we also consider 3D data at lower redshifts (z = 0, 0.5) in our investigation. The entire dataset in our study consists of products of two state-of-the-art simulations in CAMELS, IllustrisTNG\footnote{Which we will also name TNG throughout.} \citep{nelson2019illustristng} and SIMBA \citep{dave2019simba}. Each simulation produced one thousand boxes at $z = 1$ with a volume of $25\times25\times25~(h^{-1}{\rm Mpc})^3$ and a $128^3$ voxels resolution. Each 3D box corresponds to a set of cosmological ($\Omega_{\rm m}$, $\sigma_{8}$) and astrophysical parameters which are composed of the stellar feedbacks ($A_{\rm SN1}$, $A_{\rm SN2}$) and active galactic nuclei (AGN) feedbacks ($A_{\rm AGN1}$, $A_{\rm AGN2}$). Throughout our analyses, due to different prescription for the ionizing background used in each simulation, we only consider the cosmology ($\Omega_{\rm m}$, $\sigma_{8}$). It is worth noting that the instances from both simulations are from different distributions, owing to the treatment of the radiative transfer.

\begin{figure}
\includegraphics[width=0.45\textwidth]{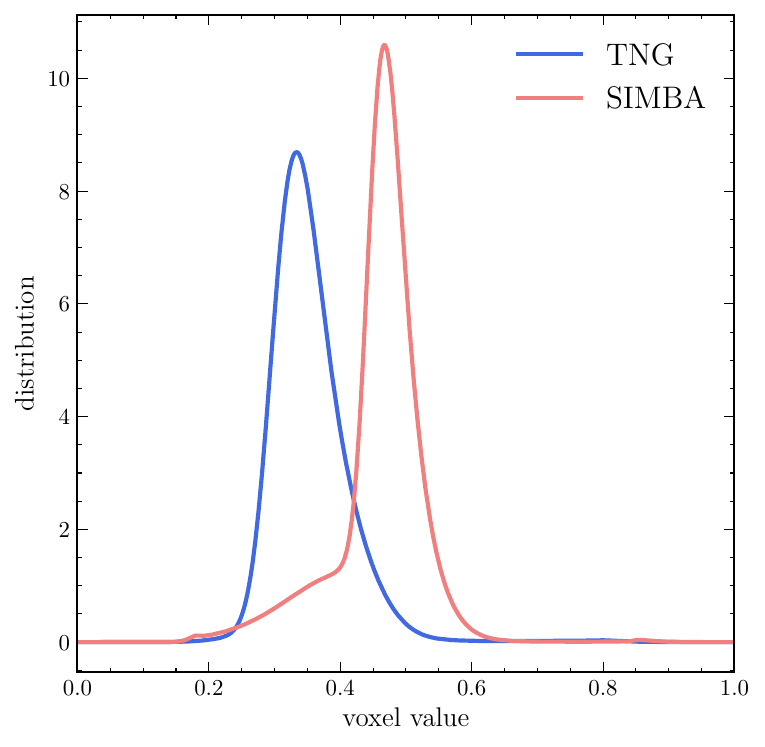}
\caption{\label{fig:voxel_dist} Probability distribution of the voxel values of the HI tomographic data from each simulation. Blue and red solid lines indicate TNG and SIMBA voxel distributions respectively. The voxel values in all 3D grids are normalized to [0,1]. It is evident that SIMBA and TNG 3D grids were generated by two different distributions.}
\end{figure}

In each dataset from one simulation (e.g. SIMBA), the 1000 instances are split into 700, 150, and 150 examples for training, validation and testing respectively. In order to mitigate overfitting, given the size of the training set, we opt for a data augmentation which mainly consists of flipping the voxels along one coordinate axis and injecting Gaussian noise. The same operations are done along the other two coordinates axes, such that in total, we have 2800 instances for training the encoder. In Figure~\ref{fig:voxel_dist} we show the probability distribution of the voxel values of both datasets, SIMBA (red line) and TNG (blue line). The voxel values of all volumes in an entire dataset are normalized to [0,1], and the voxel distribution of each instance is computed. Each plot in Figure~\ref{fig:voxel_dist} results from taking the mean of voxel distributions of all 3D grids in one dataset. There is a clear shift between the two distributions in Figure~\ref{fig:voxel_dist}, implying that one dataset is out-of-distribution with respect to the other.    

% In our approach, we first train a deep learning model on one dataset (e.g. IllustrisTNG) to map the 3D grids to cosmology.

% \begin{align}
% \Delta t \propto \left( \nu_{\text{lo}}^{-2} - \nu_{\text{hi}}^{-2} \right)\cdot\text{DM},
% \end{align}
\section{Method}\label{sec:method}
Our approach consists of two steps which are: learning to featurize the 3D boxes and using a Bayesian method to constrain the cosmological parameters. Provided the high dimensions of the boxes ($128\times128\times128$), directly retrieving the cosmology at the field level poses a challenge to the Bayesian technique. Thus, we  first need to cast the inputs into a lower dimension latent space before the probabilistic learning.

\subsection{Encoder}\label{encoder}
The dimensionality reduction of the HI 3D data is achieved via a regression task, where a deep learning network 
% composed of several three dimensional convolutional layers and two fully connected layers, 
is trained to infer the cosmology directly at the field level. The network model consists of an encoder, composed of several three dimensional convolutional layers, that learns the feature representation of the inputs, and a regressor, comprising two dense layers, that predicts the cosmological parameters. The network architecture presented in Table~\ref{encoder_arch} is structurally similar to that of the two dimensional case considered in \cite{andrianomena2023predictive}. The encoder \footnote{Which is inspired from the one in \cite{andrianomena2025towards}, a two dimensional case.} is constituted by four stages. Each stage of the first two contains two three dimensional convolutional (3DConv) layers   whereas each stage of the last ones uses three 3DConv layers. The channel number at each stage is the double of that in the previous one, whereas the dimensions (depth D $\times$ Height H $\times$ Width W) are halved using 3D \verb|Max Pooling| at the output of each stage. The regressor has two fully connected layers. To update both the encoder and regressor weights via backpropagation, we use the loss function \citep{villaescusa2021multifield, andrianomena2023predictive, andrianomena2025towards} 
\begin{table}
 \caption{The architecture of the network model in our study. "BN + ReLU" denotes a batch normalization followed by a ReLU activation. The output shape is defined by the channel C, the depth D, the height H and the width W.}
 \begin{tabular}{lll}
  \hline
   & Layer & Output shape (C, D, H, W) \\
  \hline
  \hline
  1  & Input & (1, 128, 128, 128)\\[2pt]
  2  & 3D Convolutional Layer & (16, 128, 128, 128)\\[2pt]
  3  & BN + ReLU & (16, 128, 128, 128)\\[2pt]
  4  & 3D Convolutional Layer & (16, 128, 128, 128)\\[2pt]
  5 & BN + ReLU & (16, 128, 128, 128)\\[2pt]
  6  & 3D  Max Pooling & (16, 63, 63, 63)\\[2pt]
  7  & 3D  Convolutional Layer & (32, 63, 63, 63)\\[2pt]
  8 & BN + ReLU & (32, 63, 63, 63)\\[2pt]
  9  & 3D  Convolutional Layer & (32, 63, 63, 63)\\[2pt]
  10  & BN + ReLU & (32, 63, 63, 63)\\[2pt]
  11 & 3D Max Pooling & (32, 31, 31, 31)\\[2pt]
  12 & 3D  Convolutional Layer & (64, 31, 31, 31)\\[2pt]
  13  & BN + ReLU & (64, 31, 31, 31)\\[2pt]
  14 & 3D  Convolutional Layer & (64, 31, 31, 31)\\[2pt]
  15  & BN + ReLU & (64, 31, 31, 31)\\[2pt]
  16 & 3D  Convolutional Layer & (64, 31, 31, 31)\\[2pt]
  17  & BN + ReLU & (64, 31, 31, 31)\\[2pt]
  18 & 3D Max Pooling & (64, 15, 15, 15)\\[2pt]
  19 & 3D  Convolutional Layer & (128, 15, 15, 15)\\[2pt]
  20  & BN + ReLU & (128, 15, 15, 15)\\[2pt]
  21 & 3D  Convolutional Layer & (128, 15, 15, 15)\\[2pt]
  22  & BN + ReLU & (128, 15, 15, 15)\\[2pt]
  23 & 3D  Convolutional Layer & (128, 15, 15, 15)\\[2pt]
  24  & BN + ReLU & (128, 15, 15, 15)\\[2pt]
  25 & 3D adaptive average pooling & (128, 2, 2, 2)\\[2pt]
  26 & Flattening & (1024)\\[2pt]
  27 & Dropout & (1024)\\[2pt]
  28 & Fully connected layer & (256) \\[2pt]
  29 & Dropout + Relu & (256) \\[2pt]
  30 & Fully connected layer & (4) \\[2pt]
  \hline
 \end{tabular}
 \label{encoder_arch}
\end{table}
\begin{eqnarray}\label{loss-function}
    \mathcal{L} &=& \frac{1}{N}\sum_{i = 1}^{N}{\rm log}\left(\frac{1}{M}\sum_{j = 1}^{M}(\theta_{i,j} - \mu_{i,j})^{2}\right) \nonumber\\
   && + \frac{1}{N}\sum_{i = 1}^{N}{\rm log}\left(\frac{1}{M}\sum_{j = 1}^{M}\left((\theta_{i,j} - \mu_{i,j})^{2} - \sigma_{i,j}^{2}\right)^{2}\right)
\end{eqnarray}
where $M$, $N$, $\mu_{i,j}$, $\theta_{i, j}$ and $\sigma_{i,j}$ are batch size, number of output parameters, ground truth, predictions and standard deviations related to each predicted parameter. The four outputs of the network model account for the predictions (2) and corresponding errors (2). At the end of this pre-training\footnote{A regression task whose objective is to build a map between the 3D HI data and the cosmology.}, the encoder is able extract the latent codes of the 3D inputs which are vectors of length 1024. The latter, in turn, are used to train another network within a Bayesian framework which provides the posterior uncertainties of the predicted parameters.

To build the deep regressor which is composed of a feature extractor that featurizes the high dimensional inputs, and a regressor, we choose Adam optimizer with a learning rate set to 0.0015, a batch size of 10 and a numbers of epochs equal to 200 and 150 when training on SIMBA and TNG datasets respectively. In order to have a better control over the training convergence, the learning rate is halved when the loss on the validation set does not improve over five epochs, using \verb|ReduceOnPlateau|.  
\subsection{Likelihood-free inference}
In a Bayesian analysis we have that 
% $p(\bm{\omega}|\bm{x})$
\begin{equation}\label{bayes}
    p(\bm{\omega}|\bm{x}) = \frac{p(\bm{x}|\bm{\omega})\times p(\bm{\omega})}{\int p(\bm{x}|\bm{\omega})\times p(\bm{\omega}){\rm d}\bm{\omega}},
\end{equation}
where $\bm{x}$, $\bm{\omega}$, $p(\bm{\omega}|\bm{x})$, $p(\bm{x}|\bm{\omega})$ and $p(\bm{\omega})$ are the data, the model parameters, the posterior, the likelihood and the prior respectively. The denominator in Equation~\ref{bayes} is known as the marginal likelihood which is essentially the probability of the data averaged over all values of the parameters $\bm{\omega}$. The inference is achieved by using a sampler, e.g. Monte Carlo Markov Chain (MCMC), which is predicated on the knowledge of the likelihood. 
% which, in turn, presupposes that a functional form of the model is known. 
Generally, in cosmology and astrophysics, Gaussian likelihood can be assumed and provides good estimation of the parameters of interest. However, there are cases where the likelihood is unknown, e.g. dealing with complex non-Gaussian data, or using simulation based model. In such instances, one needs to resort to a method that doesn't rely on a functional form of the likelihood, i.e. likelihood-free inference. Traditionally, Approximate Bayesian Computation \citep[ABC;][]{blum2010approximate} is one such approach which consists of simulating statistical summaries using parameters drawn from a proposal. The parameters are accepted (or rejected) depending on whether the simulated summaries are close (using a distance metric) to the observed summaries. We refer the interested reader to \cite{akeret2015approximate, jennings2017astroabc} for some applications of ABC in cosmological parameter estimation. The main issues with the ABC method (and its variants) are that it can't handle high dimensional inputs, and requires both carefully chosen summary statistics and a distance metric\footnote{This is used as a criterion to accept/reject parameters corresponding to simulated summaries.}. New approaches to likelihood-free inference involves framing the problem into a density estimation which can be broadly categorized into either likelihood or posterior estimations. The former seeks to fit a model, using simulation data \{$\bm{x}$, $\bm{\omega}$\}, to estimate the conditional density $p(\bm{x}|\bm{\omega})$ which is then fed into a sampler (e.g. MCMC) for inference; whereas the latter aims at directly estimating the posterior density $p(\bm{\omega}| \bm{x})$ using the data. For instance, 
% Synthethic likelihood method \cite[SL;][]{wood2010statistical}, 
Sequential neural likelihood \cite[SNL;][]{papamakarios2019sequential} uses Masked Autoregressive Flow \cite[MAF;][]{papamakarios2017masked} as a density estimator to approximate the likelihood. PYDELFI \citep{alsing2019fast}, a python package for DEnsity Likelihood-Free Inference, based on the prescriptions in \cite{papamakarios2019sequential}, is another method that estimates the likelihood by using Neural Density Estimator (NDE) such as Mixture Density Network \cite[MDN;][]{bishop1994mixture} or MAF. Examples of posterior estimation methods are those prescribed in \cite{papamakarios2016fast} and \cite{lueckmann2017flexible}. To improve on previous approaches, \cite{greenberg2019automatic} exploited the \textit{desirable properties} of  both likelihood (such as flexible proposals) and posterior (such as feature learning\footnote{Mapping the data to the parameters like in any regression/classification tasks.}) approximations to estimate the true posterior. In our analyses, we consider the method developed in \cite{greenberg2019automatic}, Automatic Posterior Transformation (APT).

In a simulation based model, for which the likelihood is intractable, we have data $\bm{x} \in \mathbb{R}^{D}$ \footnote{$D$ denotes the data dimension.} that are generated by a simulator which is dependent on a set of parameters 
$\bm{\omega}$. The idea consists of turning inference into a density estimation without requiring any knowledge of the likelihood $p(\bm{x}|\bm{\omega})$ and by directly building a map between the parameters $\bm{\omega}$ and the corresponding data from the simulator $\bm{x}$. The posterior density $p(\bm{\omega}|\bm{x})$ is approximated by a family of distributions $q_{\bm{\psi}}$ where $\bm{\psi}$ denotes the function parameters. In practice, $\bm{\psi}$ is derived from the data $\bm{x}$ via training a neural network $F(\bm{x}, \bm{\phi})$\footnote{In other words, $\bm{\psi} = F(\bm{x},\bm{\phi})$.}, such that $q_{F(\bm{x},\phi)}(\bm{\omega})\approx p(\bm{\omega}|\bm{x})$ \citep{greenberg2019automatic}. The parameters $\bm{\omega}$ are drawn from a prior (also known as proposal) and fed into the simulator to get the dataset \{($\bm{\omega}$, $\bm{x}$)\} which is used to build the approximator $q_{F(\bm{x},\phi)}(\bm{\omega})$. To update the weights $\bm{\phi}$ of the latter, the cross-entropy loss is used according to
\begin{equation}\label{negative_loss}
    \mathcal{L}(\bm{\phi}) = - \sum_{j = 1}^{N_{\rm samples}} {\rm log}\>q_{F(x_{j},\phi)}(\omega_{j}).
\end{equation}
It is worth noting that \cite{alsing2019fast} employed similar negative log-likelihood to build their likelihood estimator\footnote{See their Equation 7.}. Provided an observation $\bm{x}_{\rm obs}$, the network $q_{F(\bm{x}_{\rm obs},\phi)}(\bm{\omega}_{\rm obs})$, once trained, is used to estimate the posterior $p(\bm{\omega}_{\rm obs}|\bm{x}_{\rm obs})$. In our analyses, the density estimator that is selected is MAF \citep{papamakarios2017masked} which is built by stacking  four Masked Autoencoder for Distribution Estimation layers \citep[MADE;][]{germain2015made}. In a normalizing flow (NF) model \citep{rezende2015variational}, the data distribution $\bm{x}\sim p(\bm{x})$ is mapped to a base density $\bm{u}\sim \pi(\bm{u})$ using an invertible differentiable function $f$: $\bm{u}\rightarrow \bm{x}$. It is possible to approximate $p(\bm{x})$ by resorting to the change of variable formula
\begin{equation}\label{change_variable}
    p(\bm{x}) =  \pi_{\bm{u}}(f^{-1}(\bm{x})) \left|{\rm det}\left(\frac{\partial f^{-1}}{\partial \bm{x}}\right)\right|, 
\end{equation}
where $\left|{\rm det}\left(\frac{\partial f^{-1}}{\partial \bm{x}}\right)\right|$ is the determinant of the Jacobian. To estimate the data density, the latter is cast into a product of the Gaussian conditionals $p(\bm{x}) = \prod_{j} p(x_{j}|\bm{x}_{1:j-1})$ where in each component $p(x_{j}|\bm{x}_{1:j-1}) = \mathcal{N}(x_{i}| u_{i},({\rm exp}\>\alpha_{i})^{2})$ \citep{papamakarios2017masked}, the mean $u_{j}$ and variance $\alpha_{j}$ are parameterized by scalar functions, i.e. $u_{j} = f_{u_{j}}(\bm{x}_{1:j-1})$, $\alpha_{j} = f_{\alpha_{j}}(\bm{x}_{1:j-1})$), in MADE.
\begin{figure*}
\includegraphics[width=0.7\textwidth]{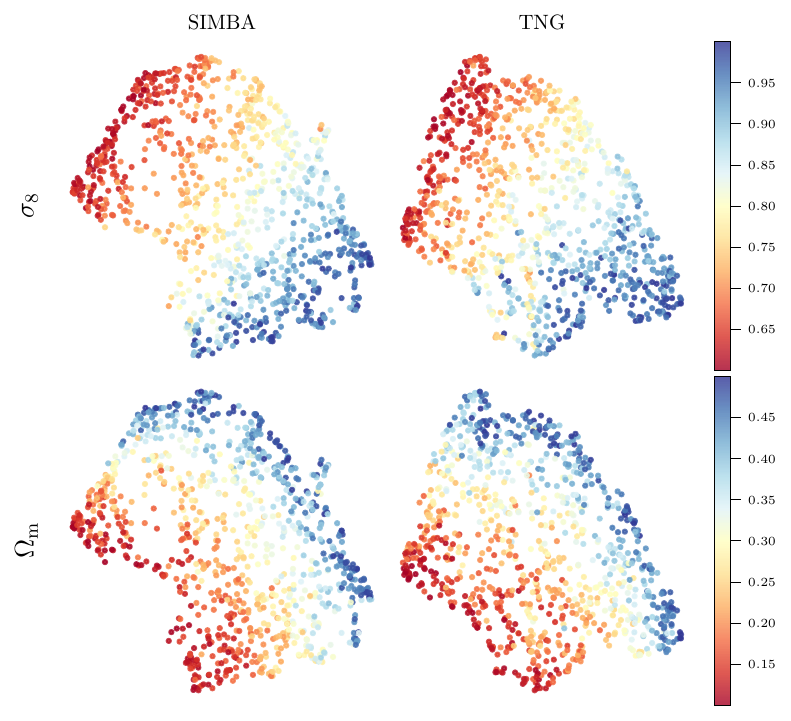}
\caption{\label{fig:embedding} The latent codes are cast into 2D space using UMAP method. The color coding of each dot in the top panels is based on the value of $\sigma_{8}$, whereas that of a dot in the bottom panels is related to $\Omega_{\rm m}$. All panels in each column correspond to a dataset, e.g. SIMBA.}
\end{figure*}
\begin{figure}
\includegraphics[width=0.48\textwidth]{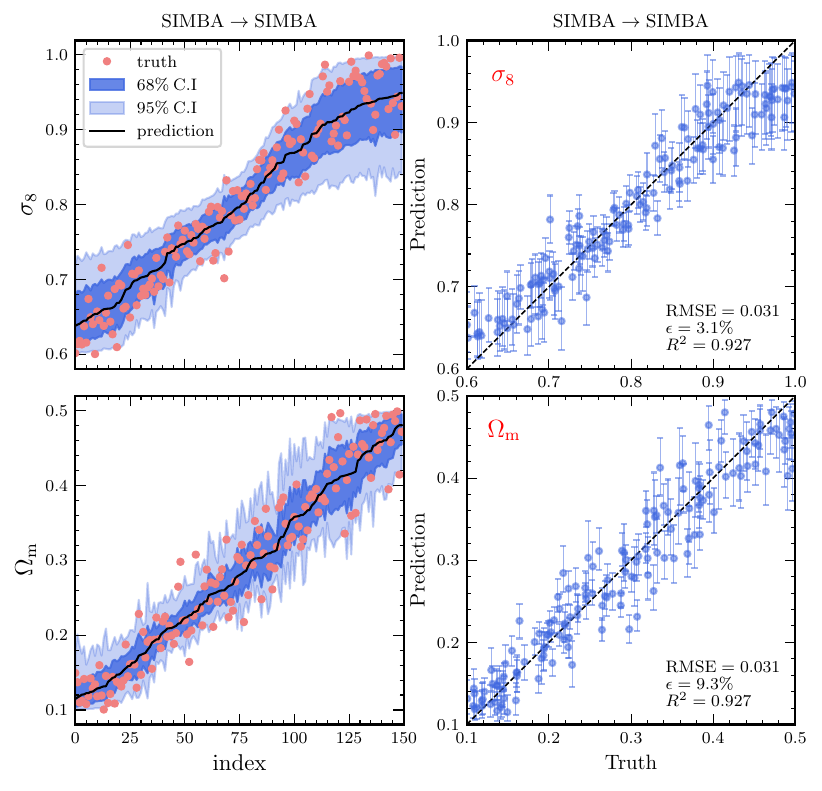}
\caption{\label{fig:posterior_prediction_simba}  
\textit{Left column}: The red dots, dark blue shaded area, light blue shaded area and black solid line indicate the ground truth, $68\%$ credible interval ($1\sigma$), $95\%$ credible interval ($2\sigma$) and the median of predictions sampled from the trained \verba{APT}. The entire HI 3D dataset from a simulation is first encoded using $E^{\rm SIMBA}$, and split into train/valid/test to train and test \verba{APT}. \textit{Right column}: Each blue dot represents a prediction from \verba{APT} whereas the error bars denote the $1\sigma$ epistemic uncertainty. The dashed black line is the identity line 1:1 in order to highlight how well the prediction correlates with the ground truth.}
\end{figure}
% Normalizing flow \citep{rezende2015variational} 
It is worth reiterating that the training dataset for the density estimator consists of the latent codes $\bm{\ell}_{c}$ (vector of length 1024) extracted by the encoder (see \S\ref{encoder}) and their corresponding cosmology, i.e. \{$\bm{\ell}_{c}$, ($\Omega_{\rm m}, \sigma_{8}$)\}. In general, it is possible to train a density estimator for more than one round within the APT framework. In the first round, during which the posterior estimator is trained for several epochs, the parameters are drawn from their prior and fed into a simulator to build the training dataset. The updated proposal for the second round is essentially the posterior that has been built in the first round (and so on). Provided that we don't have a simulator that produces data $\bm{x}$ from a given cosmology ($\Omega_{\rm m}$, $\sigma_{8}$), we use CAMELS HI 3D data for a single round training, where the proposal is the uniform prior used in the CAMELS simulation, i.e. $\Omega_{\rm m}\sim \mathcal{U}[0.1, 0.5]$ and $\sigma_8\sim \mathcal{U}[0.6, 1.0]$. We select a batch size of 20 and a learning rate of 0.0005. Our implementation makes use of the Simulation Based Inference (SBI) package \citep{tejero-cantero2020sbi,tejero2020sbi} which has a built-in early stopping such that the training ends as soon as it converges.
% 
% \cite{rezende2015variational} NF
\section{Results}\label{sec:results}
\subsection{Building the encoder}\label{subsec:encoder}
Through a regression task which uses the labels, i.e. cosmological parameters ($\Omega_{\rm m}$, $\sigma_{8}$), we train a featurizer which extracts the salient features from the 3D grids, casting the high dimensional inputs into vectors of length 1024. We fit an encoder to each dataset, in other words we have two training runs, and name each featurizer according to the dataset used to build it, i.e. $E^{\rm SIMBA}$ and $E^{\rm TNG}$ when trained on SIMBA and TNG respectively. To assess the performance of a regressor throughout our analyses, we consider coefficient of determination, $R^2$, which accounts for both the variability of the predictions and the correlation between the latter and the ground truth, according to 

\begin{equation}\label{r2score}
    R^{2} = 1 - \frac{\sum_{i = 1}^{N_{\rm samples}}\left(\theta_{i}-\mu_{i}\right)^{2}}{\sum_{i = 1}^{N_{\rm samples}}\left(\theta_{i}-\bar{\theta}\right)^{2}}.
\end{equation}
It is noted that $R^{2}$ is more stringent than Pearson correlation coefficient $\rho$ in terms of checking the performance of a regression model.

When trained on SIMBA dataset, we obtain $R^{2}$ = 0.905, 0.893 for $\Omega_{\rm m}$ and $\sigma_{8}$ respectively on the test set. It is clear that the model is able to extract both parameters from the 3D grids with similar degree of accuracy. The 2D case model in \cite{andrianomena2025towards}, which was trained and tested on SIMBA HI maps, achieved an $R^2$ = 0.72 on $\sigma_8$. Although a direct comparison can't be made between the 2D case studied in \cite{andrianomena2025towards} and this work, it can be argued that the matter distribution along the third dimension provides a meaningful information for a deep regressor model to retrieve the amplitude of the density fluctuations. The model, trained and tested on TNG datasets, is able to constrain the cosmology with an accuracy of $R^2$ = 0.926, 0.949 on $\Omega_{\rm m}$ and $\sigma_{8}$ respectively. The deep network is able to constrain both cosmological parameters to a  good accuracy overall, but similar to the $\sigma_8$ constraint obtained from SIMBA 3D grids, it appears that extracting $\sigma_8$ from TNG 3D grids is more precise than that from TNG 2D maps. 

The predictive power of the deep regressor is predicated on the encoder that is capable of projecting the high dimensional inputs onto a separable latent space. We reiterate that the main objective of this first step is to fit the encoder such that it learns a lower dimensional features of the 3D dataset. We present in Figure~\ref{fig:embedding} the representations that have been learned by the encoder in both SIMBA and TNG cases. For visualization purposes, the learned features, which are vectors of length 1024 at the output of an encoder ($E^{\rm SIMBA}$ or $E^{\rm TNG}$), are further projected onto two dimensional space using UMAP method \citep{mcinnes2018umap}. To arrive at the results in the first and second columns of Figure~\ref{fig:embedding}, UMAP models are trained on the outputs of $E^{\rm SIMBA}$ and $E^{\rm TNG}$ respectively. The color codings on the top and bottom panels are based on the values of $\sigma_8$ and $\Omega_{\rm m}$ respectively. In all panels, the dots\footnote{They represent instances.} with similar cosmological parameter values are well clustered in this feature space, indicative of the separability of the learned representations. This accounts for the good performance of the deep regressors overall. Therefore, it can inferred from Figure~\ref{fig:embedding} that both encoders $E^{\rm SIMBA}$ and $E^{\rm TNG}$ are capable of summarizing the 3D grids. In the following section, for each dataset, we make use of the latent codes $\bm{\ell}_{c}$ from the encoder to build a posterior estimator, i.e. $p(\bm{\omega}|\bm{\ell_c})$.
\begin{figure}
\includegraphics[width=0.48\textwidth]{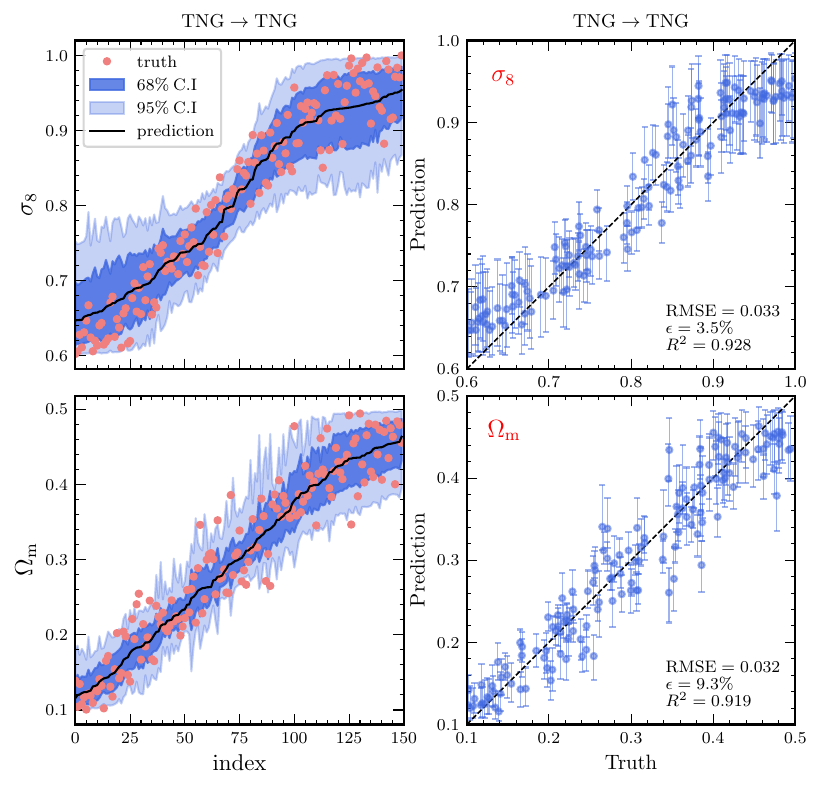}
\caption{\label{fig:posterior_prediction_tng} Similar to the results presented in Figure~\ref{fig:posterior_prediction_simba} for the TNG$\rightarrow$TNG where the \verba{APT} is trained on the TNG representations extracted by $E^{\rm TNG}$.}
\end{figure}

\begin{figure}
\includegraphics[width=0.47\textwidth]{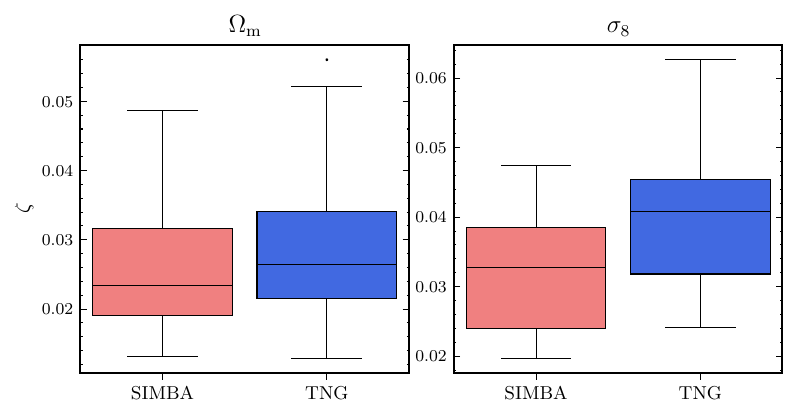}
\caption{\label{fig:box_uncertainty} Distribution of the posterior uncertainty related to each parameter in each setup. Red and blue boxes correspond to SIMBA and TNG datasets respectively.}
\end{figure}

\begin{figure}
\includegraphics[width=0.48\textwidth]{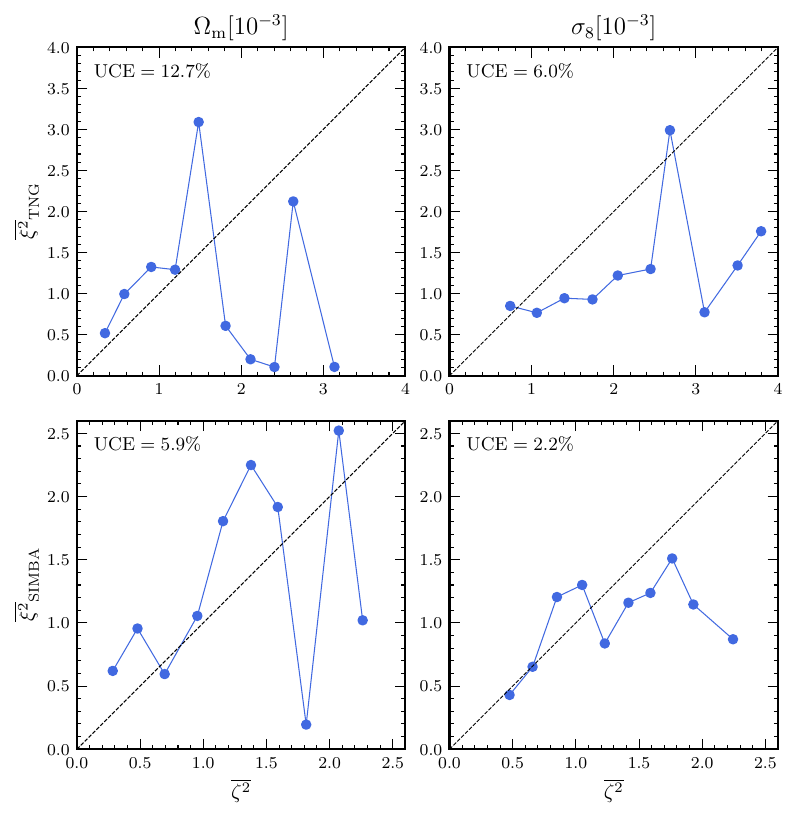}
\caption{\label{fig:calibration_uncertainty} Each panel shows the variation of squared error as a function of the squared of epistemic uncertainty, indicated by each blue dot. The black dotted line denotes the identity line where the squared error is equal to the epistemic variance. For better visualization, both the error and uncertainty are multiplied by $10^3$. The Uncertainty Calibration Error (UCE) for each parameter in each case is shown on the top left of each panel.}
\end{figure}

\begin{figure}
\includegraphics[width=0.48\textwidth]{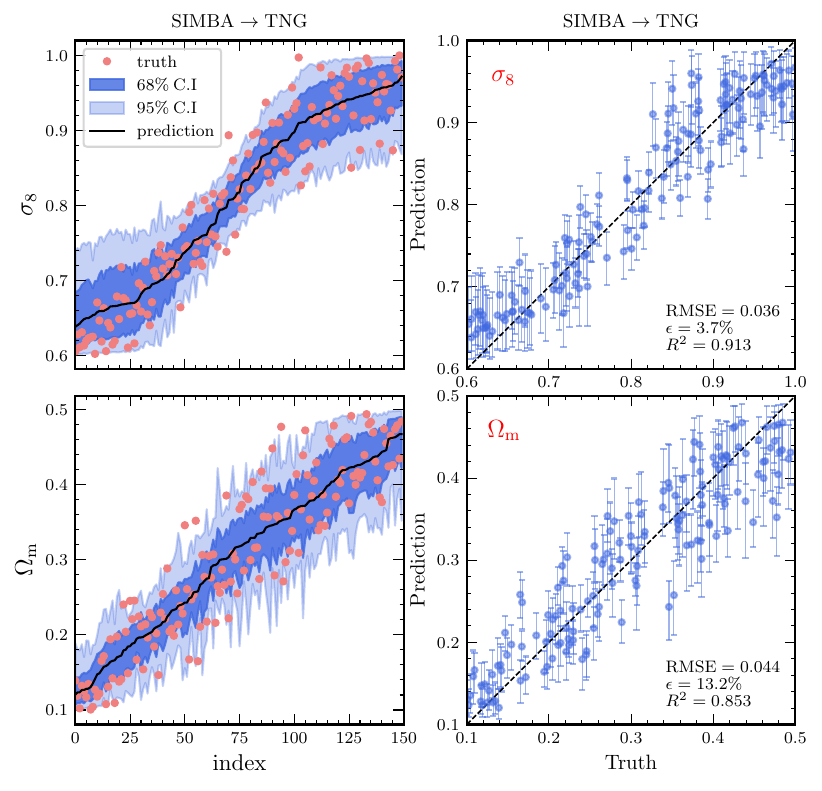}
\caption{\label{fig:posterior_prediction_simba_tng}Similar to Figures~\ref{fig:posterior_prediction_simba} and \ref{fig:posterior_prediction_tng} but \verba{APT} method is trained on TNG features encoded by $E^{\rm SIMBA}$.}
\end{figure}
\begin{figure}
\includegraphics[width=0.48\textwidth]{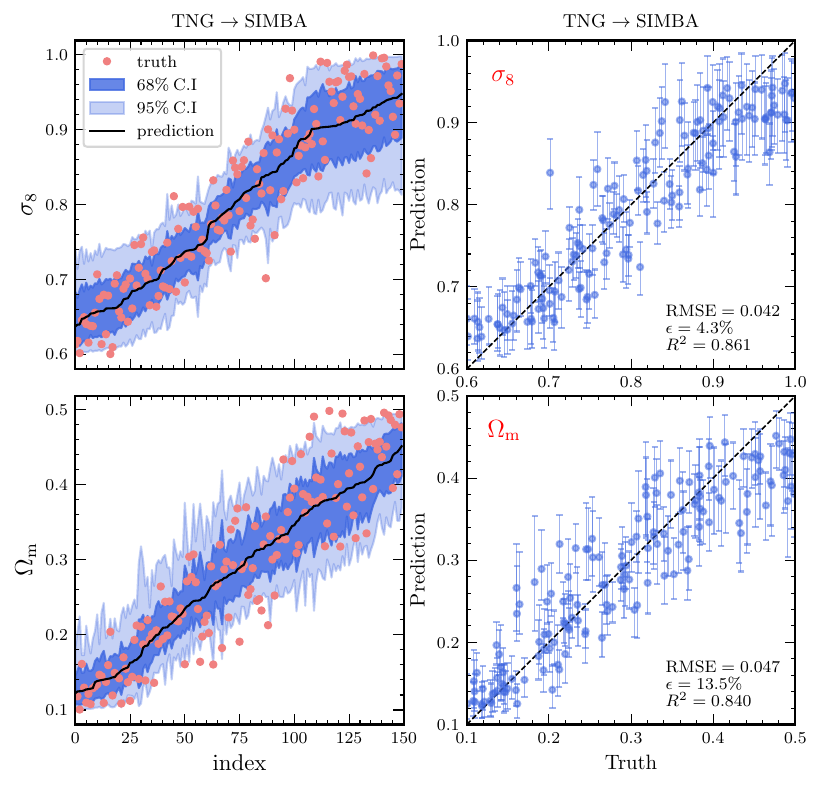}
\caption{\label{fig:posterior_prediction_tng_simba} Similar to Figures~\ref{fig:posterior_prediction_simba}, \ref{fig:posterior_prediction_tng} and \ref{fig:posterior_prediction_simba_tng} but \verba{APT} method is trained on SIMBA features encoded by $E^{\rm TNG}$.}
\end{figure}
\subsection{Fitting the posterior estimator}
For easy reference, we first provide the notations for all the setups considered in our analyses. \verba{Dataset}$_1$$\rightarrow$\verba{Dataset}$_2$ denotes that the density estimator is trained/tested on the 3D grid summaries of \verba{Dataset}$_2$ that are extracted by an encoder built from \verba{Dataset}$_1$, e.g. SIMBA$\rightarrow$SIMBA indicates that the SIMBA learned representations, which are used to train the neural density estimator, are the outputs of $E^{\rm SIMBA}$. We first focus on two cases, SIMBA$\rightarrow$SIMBA and TNG$\rightarrow$TNG.
\begin{table*}%[!ht]
\caption{For ease of reference, the notation \verba{Dataset}$_1$$\rightarrow$\verba{Dataset}$_2$ indicates that the encoder has been trained on \verba{Dataset}$_1$ and is used to extract the representations of \verba{Dataset}$_2$ that are in turn used to build various regressors, e.g. \verba{APT}. The values presented are $R^{2}$ score achieved on $\Omega_{\rm m}$ and $\sigma_{8}$ using the regressors in a given setup.}
  % \label{tab:uncertainty}
  \centering
  \begin{tabular}{| *{9}{c|} }%{|c|c|c|c|}
    %\toprule
    % \cmidrule(r){1-2}
    \hline
     % & SN(sd)& $\rm TN^{ADDA}$(td) &  $\rm TN^{OT}$(td)\\
    & \multicolumn{2}{c|}{\verba{APT}} & \multicolumn{2}{c|}{\verba{GP}} & \multicolumn{2}{c|}{\verba{RF}} & \multicolumn{2}{c|}{\verba{LIN}} \\\cline{2-9}
    %\midrule
    
    & $\Omega_{\rm m}$ & $\sigma_{8}$ & $\Omega_{\rm m}$ & $\sigma_{8}$ & $\Omega_{\rm m}$ & $\sigma_{8}$ &  $\Omega_{\rm m}$ & $\sigma_{8}$   \\
    \hline
    SIMBA$\rightarrow$SIMBA & \textbf{0.927} & \textbf{0.927} & 0.893 & 0.798 & 0.835 & 0.923 &  0.874 & 0.935   \\
    \hline
    SIMBA$\rightarrow$TNG & \textbf{0.853} & \textbf{0.913} & 0.807 & 0.790 & 0.824 & 0.918 & 0.801 & 0.883   \\
    %\bottomrule
    \hline
    TNG$\rightarrow$TNG & \textbf{0.919} & \textbf{0.928} & 0.893 & 0.904 & 0.840 & 0.920 & 0.891 & 0.917  \\
    %\bottomrule
    \hline
    TNG$\rightarrow$SIMBA & \textbf{0.840} & \textbf{0.861} & 0.809 & 0.731 & 0.765 & 0.824 & 0.801 & 0.806   \\
    %\bottomrule
    \hline
  \end{tabular}
\label{tab:rsquare}
\end{table*}
At the end of its training, the neural density estimator is tested on a held-out set. For each prediction, we draw from the posterior $p(\mu|\ell^{\rm test^*}_{c})$ -- where $\ell^{\rm test^*}_{c}$ is a single instance from the test set -- 10,000 samples from which we compute the median (50$^{\rm th}$ percentile), and the predictive uncertainties, i.e. 68$\%$  and 95$\%$ credible intervals. For SIMBA$\rightarrow$SIMBA setup, the results are shown in the left panels of Figure~\ref{fig:posterior_prediction_simba}. In what follows, predicted value of a parameter means its median value from the 10,000 samples drawn from the posterior. It is worth noting that, to arrive at the plot on the left panels of Figure~\ref{fig:posterior_prediction_simba}, the labels, predictions and uncertainties are all sorted in ascending order (according to the predictions) such that the x-axis of each panel is the array index whereas the y-axis is the corresponding parameter value. Each red dot indicates the ground truth of a parameter, the solid black line is the predictions whereas the dark and light shaded blue areas are the 68$\%$ ($1\sigma$) and 95$\%$ ($2\sigma$) CI respectively. In general, most of the true values of the cosmological parameters fall within $1\sigma$, although there are some outliers that are even beyond two standard deviations. It can be noticed that for both $\Omega_{\rm m}$ and $\sigma_{8}$ the uncertainties corresponding to high parameter values are relatively larger, which can be accounted for by lower number of instances in those ranges in the training dataset, inducing a bias. The posterior uncertainty ($\zeta$), which is also known as epistemic uncertainty, reflects what the model doesn't know about the data. Hence, in a lower data regime, the model uncertainty is expected to be larger. To assess how strong the correlation between the ground truth and its corresponding prediction is, we plot the latter as a function of the former in the right columns of Figure~\ref{fig:posterior_prediction_simba}. Each blue dot, the error bars and the dashed black line indicate the prediction, $1\sigma$ uncertainty and identity line respectively. In each panel on the right column, we show the root mean-squared error ${\rm RMSE} = \sqrt{\frac{1}{N_{\rm samples}}\sum_{i=1}^{N_{\rm samples}}(\theta_{i} - \mu_{i})^2}$, the mean relative deviation $\epsilon = \frac{1}{N_{\rm samples}}\sum_{i = 1}^{N_{\rm samples}}\frac{|\theta_{i} - \mu_{i}|}{\mu_{i}}$, and the coefficient of determination $R^2$. The results suggest that \verba{APT} method achieves cosmological constraints that are tighter than those obtained from the deep regressor in \S\ref{subsec:encoder}, i.e. the coefficient of determination obtained with \verba{APT} for each parameter is $R^{2}\ge 0.92$. The mean relative deviation on $\sigma_{8}$ (see Figure~\ref{fig:posterior_prediction_simba} top righ panel) is on a par with that achieved in \cite{andrianomena2023predictive} from using HI 2D maps, however results show that the matter density exhibits slightly larger deviation on average compared to the 2D case (see bottom right panel of Figure 3 in \cite{andrianomena2023predictive}). 

In the TNG$\rightarrow$TNG setup, it seems that the \verba{APT} model ignorance is also more pronounced at higher parameter value range, and that bias is even visible around the limits of $\sigma_{8}$ prior range, as can be seen in the left column of  Figure~\ref{fig:posterior_prediction_tng}. Neverthless, as also observed in the SIMBA$\rightarrow$SIMBA case, most of the true parameter values are within $1\sigma$ uncertainty, despite some outliers. The performance of the posterior estimator is consistent with that of SIMBA$\rightarrow$SIMBA case as evidenced by similar values of the obtained RMSE, $\epsilon$ and $R^2$ for all parameters (see panels in the right column of Figure~\ref{fig:posterior_prediction_tng}). Although its performance is promising in general, i.e. $R^2 \ge 0.91$ for all parameters, the \verba{APT} method is slightly outperformed by the deep regressor (in \S\ref{subsec:encoder}) on TNG$\rightarrow$TNG case. 

\subsection{Posterior uncertainty}
One of the main advantages of using a probabilistic model is its ability to estimate the uncertainty inherent to its prediction. This provides some insights into the model limitation, e.g. what it doesn't know. Summary plots of the uncertainty distribution obtained for each parameter in each scenario (SIMBA$\rightarrow$SIMBA and TNG$\rightarrow$TNG) are shown in Figure~\ref{fig:box_uncertainty}. Given the asymmetric nature of the posterior uncertainty which is estimated from using the percentiles (i.e. the upper and lower uncertainties are different), we only select the upper uncertainty\footnote{Lower uncertainty can also be chosen but for illustration purposes, we only select the upper one.} to arrive at Figure~\ref{fig:box_uncertainty} and in what follows. The box represents the interquartile range (IQR), whereas its whiskers correspond to minimum/maximum. Red and blue boxes are related to uncertainty distribution of predictions in SIMBA and TNG datasets respectively. It is noticed that the median\footnote{The horizontal black solid line in a box.} obtained from SIMBA is slightly lower than that in TNG for all parameters, and the distribution support, defined by the minimum and maximum, related to SIMBA is also narrower overall. However, for each parameter, the IQRs corresponding to both datasets are consistent with each other. 

In a real world scenario, there is no ground truth to compare a deep learning model prediction with. Therefore, once a probabilistic model has been trained on simulated data, we need to assess whether its predictive uncertainty is calibrated. In other words, one should ensure that squared errors $\xi^2$ (obtained during testing) are consistent with the estimated uncertainties. For instance, in the case of variational inference using Monte Carlo Dropout the produced uncertainty is likely to be miscalibrated \citep{laves2020well}, such that it is either underestimated or overestimated. In that case, the uncertainty needs to be calibrated in order to be more reliable. In our investigation, we follow the approach in \cite{laves2020well}, named Uncertainty Calibration Error (UCE), to assess how well calibrated the epistemic uncertainty produced by our probabilistic model is. We have that \citep{laves2020well}
\begin{equation}\label{eq:uce}
    {\rm UCE} = \sum_{m = 1}^{N_{\rm bins}}\frac{|B_{m}|}{N_{\rm samples}}|\overline{{\xi}^2}(B_{m}) - \overline{\zeta^2}(B_{m})|,
\end{equation}
where $N_{\rm bins}$ is the number of bins\footnote{The uncertainty values obtained from the test set are binned.}, which is set to 10, following \cite{andrianomena2023predictive}, $B_{m}$ is the set of data points in a given bin and $|B_{m}|$ the cardinal number; the average squared error $\overline{\xi^2}(B_{m})$ in each bin is
\begin{equation}\label{eq:average-error}
    \overline{\xi^2}(B_{m}) = \frac{1}{|B_{m}|}\sum_{i \in B_{m}} ||\mu_{i} - \theta_{i}||^{2},
\end{equation}
and the average epistemic variance within a bin 
\begin{equation}\label{eq:epistemi-variance}
    \overline{\zeta^2}(B_{m}) = \frac{1}{|B_{m}|}\sum_{i \in B_{m}} \zeta^2_{i}.
\end{equation}
In order to compare the miscalibration (UCE value in $\%$) corresponding to one parameter prediction with that corresponding to another, we opt for the same modified version of UCE in \cite{andrianomena2023predictive}, according to 
\begin{equation}\label{eq:uce-modified}
    {\rm UCE_{modified}}[\%] = 100\times\frac{1}{N_{\rm bins}}\sum_{m = 1}^{N_{\rm bins}}\frac{|B_{m}|}{N_{\rm samples}}\left|\frac{\overline{{\xi}^2}(B_{m}) - \overline{\zeta^2}(B_{m})}{\overline{{\xi}^2}(B_{m})}\right|.
\end{equation}
The results are presented in Figure~\ref{fig:calibration_uncertainty}. Each panel shows the variation of the mean square error $\overline{\xi^{2}}$ as a function of mean epistemic variance $\overline{\zeta^2}$ in each bin. The UCE related to each plot is shown on the top left corner of each panel in Figure~\ref{fig:calibration_uncertainty}. For better visualization, all numbers are scaled by multiplying them by $10^3$. In an ideal case, a perfectly calibrated uncertainty, the blue dots lie along the identity line, which translates to UCE = $0\%$, in other words $\xi^{2}$ and $\zeta^2$ are perfectly correlated. It is worth noting that when a blue dot is above the 1:1 line, it means that the uncertainty is underestimated, otherwise (blue dot below 1:1 line) it is overestimated. Generally, we find larger deviation of $\zeta^2$ from $\xi^2$ at higher value of the uncertainty. Results suggest that the uncertainty obtained from SIMBA is better calibrated than the one from TNG, as evidenced by the UCE values. It also appears that a slightly better uncertainty calibration is achieved on $\sigma_8$ predictions, compared to $\Omega_{\rm m}$. An overall deviation at a percent level suggests that the posterior uncertainty is reasonably calibrated. Unlike the case studied in \cite{andrianomena2023predictive} where MC dropout method, which was used to estimate the predictive uncertainty from their model, was prone to miscalibration such that to mitigate the latter, they resorted to a method named $\sigma$ scaling \citep{laves2020well}. The predictive uncertainty in \cite{andrianomena2023predictive} includes both epistemic and aleatoric uncertainties which are added in quadrature to compute the predictive uncertainty. In the present work, we don't model the aleatoric uncertainty, which takes into account the intrinsic noise in the data, therefore there can't be a fair comparison between our analyses and those in \cite{andrianomena2023predictive} in terms of uncertainty estimation. The main idea behind computing the UCE for each parameter in each dataset is to establish whether the posterior uncertainty from our probabilistic regressor is miscalibrated.

\subsection{Testing the robustness of the learned representations}
% The main advantage of utilizing Bayesian approach within the context of parameter inference is its ability to provide predictive uncertainty, 
We have seen previously that the latent codes are such that they can be utilized to constrain cosmology to a very promising accuracy, as indicated by the resulting coefficient of determination achieved by the density estimator in all setups (i.e. SIMBA$\rightarrow$SIMBA and TNG$\rightarrow$TNG). In this section, we further investigate how meaningful and robust the learned representations are. To this end, we train the posterior estimator on latent codes of a dataset that are retrieved by an encoder built using a different dataset, in other words the cases: SIMBA$\rightarrow$TNG and TNG$\rightarrow$SIMBA where $E^{\rm SIMBA}$ and $E^{\rm TNG}$ are used to featurize TNG and SIMBA respectively.

It is apparent from the left column of Figure~\ref{fig:posterior_prediction_simba_tng} that, for the SIMBA$\rightarrow$TNG scenario, the predictive uncertainties on the parameters degrade, i.e. the credible intervals are larger than those obtained in the SIMBA$\rightarrow$SIMBA case (see Figure~\ref{fig:posterior_prediction_simba}). This is expected since $E^{\rm SIMBA}$ is attempting to compress out-of-distribution examples from TNG. Despite that, the predictive power of the neural density estimator is such that it still achieves an $R^2 \ge 0.85$ overall. Interestingly, the prediction on the matter density parameter is the most impacted, reaching $R^2 = 0.853$ which translates to an average deviation of $\epsilon = 13.2\%$, an increase of about $4\%$ with respect to the SIMBA$\rightarrow$SIMBA case. The constraint on $\sigma_{8}$ on the other hand is consistent with that obtained in SIMBA$\rightarrow$SIMBA scenario. This can potentially be explained by the fact that the encoder $E^{\rm SIMBA}$, although being fed with out-of-distribution samples, is still able to extract the salient features related to the clustering of matter.
\begin{figure}
\includegraphics[width=0.48\textwidth]{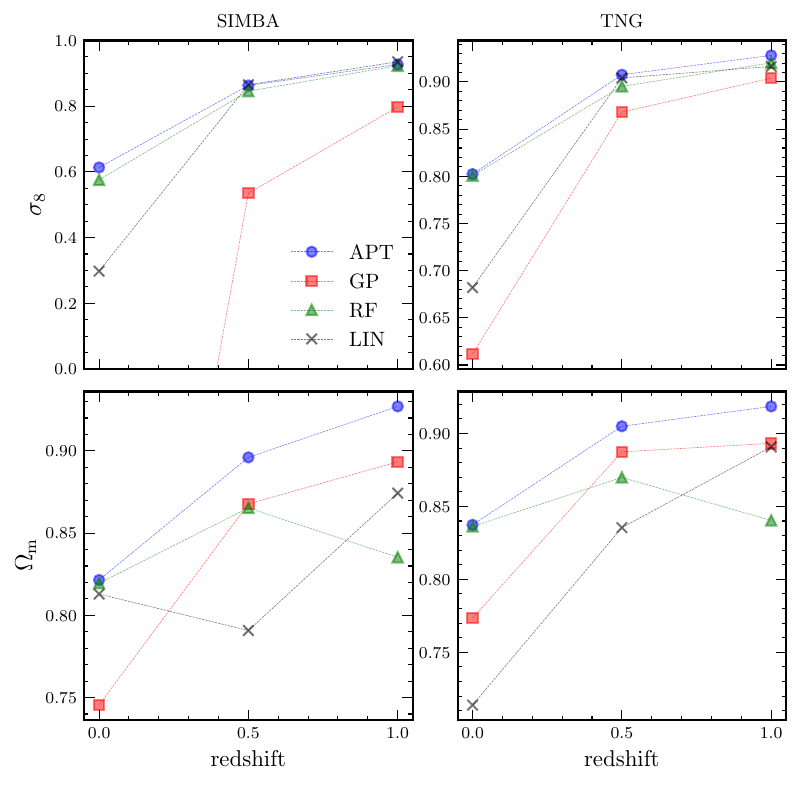}
\caption{\label{fig:prediction_algos_redshift_tng_simba} $R^2$ obtained on each cosmological parameter for each algorithm as a function of redshift. Blue dot, red square, green triangle and black cross denote $R^2$ value achieved by \verba{APT}, \verba{GP}, \verba{RF} and \verba{LIN} methods respectively. For all results corresponding to SIMBA and TNG datasets, the encoders used to extract the representations of the 3D grids are $E^{\rm SIMBA}$ and $E^{\rm TNG}$ at $z = 1.0$ respectively. The $R^2$ achieved by \verba{GP} at $z = 0$ is not shown on the top left panel as the method fails to predict $\sigma_{8}$ ($R^2 < 0$) in SIMBA at that redshift.}
\end{figure}
In a similar way, the overall epistemic uncertainty produced by the density estimator in TNG$\rightarrow$SIMBA case worsens in comparison with that estimated in TNG$\rightarrow$TNG. This is shown in the left column in Figure~\ref{fig:posterior_prediction_tng_simba}. The correlation strengths between the prediction and ground truth for both cosmological parameters seem to be similar, as evidenced by $R^2$ = 0.840, 0.861 related to $\Omega_{\rm m}$ and $\sigma_{8}$ respectively. Unlike the SIMBA$\rightarrow$TNG case, building the \verba{APT} with SIMBA latent codes, whose corresponding 3D grids are seen as out-of-distribution by the encoder $E^{\rm TNG}$ which outputs them (TNG$\rightarrow$SIMBA case), negatively impacts its performance on predicting both parameters. The percent level increase of the mean relative deviation of the matter density prediction ($4\%$), as opposed to the case of the amplitude of density fluctuations whose $\epsilon$ worsens at only a sub-percent level ($0.8\%$), implies that $\Omega_{\rm m}$ prediction is the most affected. In general, we find that the parameter inference in TNG$\rightarrow$SIMBA setup is slightly more challenging than that in SIMBA$\rightarrow$TNG. This is reminiscent of the trend observed in \cite{andrianomena2025towards} where  the asymmetric nature of extracting information from out-of-distribution samples was highlighted. They found that training a network model on dataset with better resolution (TNG HI 2D maps), i.e. more details/information on high frequency scales, and predicting cosmology from out-of-distribution maps where small scale features are a bit blurred (SIMBA HI 2D maps) proved to be more challenging than the other way round. In \cite{andrianomena2025cosmological}, they also demonstrated a better generalization of the network when trained on (fake) maps produced by a generative model, noisy ones, in order to infer cosmology from TNG 2D maps which were considered to be clean ones. It is apparent from Figure~\ref{fig:voxel_dist} that the probability distribution of the voxel values of SIMBA dataset (red line) is more complex than the TNG counterpart, therefore extracting the salient features from SIMBA 3D grids is slightly more involved, similar to the 2D case studied in \cite{andrianomena2025towards}.

Overall the results obtained from the two test scenarios suggest that pre-trained encoders are able to produce meaningful feature representations (of the high dimensional inputs) in such a way that the posterior estimators trained on the latter are still capable of inferring the cosmology to a reasonable accuracy, as evidenced by $R^2 \ge 0.84$ in all cases selected. 

\subsection{Predictive power of the neural density estimator}
The idea is to assess how powerful the technique considered is, within the specific context elaborated in this work, compared to other methods. We then select three other algorithms that will be trained on the featurized 3D grid data, and test their performance in the four scenarios that have been looked at in previous sections. In our investigation, we have Gaussian process \citep[\verba{GP};][]{williams2006gaussian} whose kernel is set to be a radial basis function, Random Forest \citep[\verba{RF};][]{ho1995random} with 100 estimators, and a neural network composed of a single dense layer with 1024 units\footnote{Same as the vector length of the output from the encoder.}. The latter is known as linear test (\verba{LIN}) which simply consists of training the single dense layer on the representations for 100 epochs without using any activation function. 

We present in Table~\ref{tab:rsquare} the results obtained in all scenarios for each of the methods selected. For ease of reference, the resulting $R^2$ obtained from \verba{APT} are also included in the table. Overall, it is clear that the posterior estimator outperforms the other algorithms in terms of inferring cosmology from the latent codes. \verba{RF} method, known to be powerful as well when dealing with lower dimensional input features, seems to exhibit a similar performance on retrieving the amplitude of the density fluctuations, but its ability to extract the matter density is outperformed by \verba{APT} in general. 
% \verba{RF} method tends to be more accurate when retrieving $\sigma_8$, compared to predicting $\Omega_{\rm m}$.
% Inferring $\sigma_8$ in the setup SIMBA$\rightarrow$TNG proves to be challenging for \verba{LIN}, i.e. $R^2$ = 0.472, otherwise the 
The prediction accuracy of \verba{LIN} in all setups for both cosmological parameters is $R^2 \ge 0.80$ in general. 
This is indicative of the fact that, the learned representations, even in the case where the 3D grids are out-of-distribution to the encoder, still carries meaningful cosmological information. \verba{GP} method, which is able to reasonably predict the cosmology from the latent codes, tends to constrain $\Omega_{\rm m}$ better than $\sigma_8$. The ability of other algorithms to infer the parameters points to the fact that the performance of the neural density estimator is not solely due to its predictive power, but also thanks to the quality of the representations learned by the encoders.  

% \begin{itemize}
%     \item Gaussian process \citep[\verba{GP};][]{williams2006gaussian}, a non-parametric model which assumes that a set of real-valued function $\bm{f}$ at several points $\bm{x}$ is described by a joint Gaussian distribution. The algorithm is defined by the mean $\bm{\xi}(\bm{x})$ and the covariance $\bm{K}(\bm{x}, \bm{x}')$, where the kernel $\bm{K}$ is chosen to be a radial basis function.
%     \item \cite{ho1995random}
% \end{itemize}

\subsection{Encoding out-of-distribution boxes at different redshifts}
We further investigate the robustness of the encoded features by training/testing \verba{APT}, \verba{GP}, \verba{RF} and \verba{LIN} on latent codes of 3D boxes from different redshifts. Thus, given a simulation (e.g. SIMBA), its HI tomographic data at $z = 0$ are first compressed by an encoder, which has been trained on dataset at $z = 1$ from the same simulation (e.g. SIMBA as well), and utilized to train the four algorithms. The 3D boxes at $z = 0.5$ are processed in a similar way. 
% It is noted that we carry out the investigation, considering both simulations (SIMBA and TNG).

The results are shown in Figure~\ref{fig:prediction_algos_redshift_tng_simba} where each coefficient of determination achieved by each method at each redshift is plotted in each panel. The panels on the left column correspond to the performance of the methods when using datasets from SIMBA whereas those on the right are related to TNG. The panels on top and bottom rows show constraints on $\sigma_8$ and $\Omega_{\rm m}$ respectively. The blue dot, red square, green triangle and black cross markers are the $R^2$ values obtained from \verba{APT}, \verba{GP}, \verba{RF} and \verba{LIN} respectively. On SIMBA datasets, the general trend exhibited by all algorithms is that the prediction accuracy decreases as we go to lower redshifts ($z$ = 0 and 0.5). This can be accounted for by the fact that the quality of the 3D grid representations extracted by $E^{\rm SIMBA}$ degrades, owing to galaxy evolution (e.g. merging, ram pressure stripping etc) which impacts on the HI distribution. Nevertheless, \verba{RF} method performance doesn't appear to be sensitive to the redshift information when predicting $\Omega_{\rm m}$, showing a small peak at $z = 0.5$ (see left bottom panel in Figure~\ref{fig:prediction_algos_redshift_tng_simba}). We find that \verba{GP} fails to predict $\sigma_8$ at $z = 0$ in SIMBA, and is not therefore shown in top left panel. In the worst case scenario at $z = 0$, \verba{APT} still boasts an $R^2\sim 0.6$ on $\sigma_{8}$. On TNG datasets, we observe similar trend where the constraint on a parameter worsens as we go to lower redshifts. Consistent with results in SIMBA, \verba{RF} seems to be also less impacted (compared to other methods) by the redshift information when inferring the matter density. Interestingly, at present time (i.e. $z = 0$), all methods put tighter constraint on $\Omega_{\rm m}$ than on $\sigma_8$. This is expected since the latter is more dependent on the HI clustering which evolves with time. 
% as more neutral gas is consumed in order to form more galaxies at present. 
In general, regardless of the simulation considered and the parameter to be extracted, the neural density estimator consistently outperforms the other methods at each redshift. We also find that \verba{APT} still shows very promising performance when trained on 3D grid representations at $z\sim 0.5$ that are extracted by encoder built on dataset at $z = 1$, as evidenced by $R^2 \ge 0.83$ on all cosmological parameters in all simulation datasets considered.

\section{Conclusions}\label{sec:conclusion}
We have demonstrated how the cosmology ($\Omega_{\rm m}$, $\sigma_{8}$) can be probabilistically inferred from 3D tomographic HI data at intermediate redshift. We use both SIMBA and TNG HI 3D datasets from the CAMELS Project. Our approach consists of two steps which involve training an encoder to learn robust representations of the 3D boxes via a regression task and training a neural density estimator on the latent codes that are output by the encoder. In the first step, a deep encoder, composed of several three dimensional convolutional layers, has been trained simultaneously with a regressor, composed of two dense layers, to predict the matter density and the amplitude of density fluctuations. In total, there are two runs to train the encoders $E^{\rm SIMBA}$ and $E^{\rm TNG}$ on SIMBA and TNG datasets respectively.
% an encoder is built via a regression task. 
Once an encoder related to a given dataset has learned the representations of the latter, it is utilized to extract the entire dataset latent codes which are, in turn, used to train/test a posterior estimator in a likelihood-free inference setup. We make use of Automatic Posterior Transformation (\verba{APT}), a method that exploits the capacity of a Masked Autoregressive Flow (MAF) to estimate density, to learn the posterior. At inference time, provided a datum, which is an unseen 3D box representation from a test set, we generate 10,000 parameter samples from the posterior. The epistemic uncertainties, $1\sigma$ and $2\sigma$ errors, are computed using the samples from the posterior distribution, and the median is considered as the prediction. We investigate how well calibrated the epistemic uncertainty from the \verba{ATP} model is, by computing the Uncertainty Calibration Error (UCE). To assess both the robustness of the encoded features and the predictive power of the \verba{APT}, we consider scenarios whereby the posterior esimator is trained/tested on latent codes of dataset which is out-of-distribution with respect to the pre-trained encoder that extracts them. In our analyses, the performance of the \verba{APT} method is compared to that of other techniques, namely Gaussian Process (\verba{GP}), Random Forest (\verba{RF}) and a single dense layer network as linear test (\verba{LIN}). Furthermore, we investigate whether it is possible to recover the cosmology from features of HI 3D grid dataset at different redshifts ($z = 0, 0.5$) that are produced by an encoder built by using the same simulation dataset but at redshift $z = 1$.

Using coefficient of determination ($R^2$) as a metric to assess the performance of the deep network, composed of the encoder and regressor, we find that in each 3D dataset considered (either SIMBA or TNG), the deep regressor is able to constrain both cosmological parameters to relatively good accuracy, i.e $R^2 \ge 0.89$. In comparison with the inference on HI maps studied in \cite{andrianomena2025towards}, it appears that the 3D network in this work predicts $\sigma_8$ with a better precision, which, among other factors, can be attributed to the information from the third dimension of the training data. The results on the regression task (i.e. inferring cosmology) are corroborated by the fact that the learned representations are separable in a lower dimension latent space. In other words, data points corresponding to similar values of $\Omega_{\rm m}$ and $\sigma_8$ are well clustered in feature space (see Figure~\ref{fig:embedding}).

In both cases, training/testing \verba{APT} on SIMBA and TNG latent codes extracted by the encoders from SIMBA and TNG respectively, results suggest that most of the true values of the cosmological parameters are within $1\sigma$ uncertainty of the predicted ones (see Figures~\ref{fig:posterior_prediction_simba} and \ref{fig:posterior_prediction_tng} for SIMBA and TNG respectively). The performance of the neural density estimator is comparable to that of the 3DCNN model on TNG dataset, whereas on SIMBA dataset, the \verba{APT} marginally improves on the results achieved by the deep regressor.

By using the upper uncertainty to investigate miscalibration, we find that on SIMBA, the \verba{APT} model produces posterior uncertainty that is slightly better calibrated, in comparison with that resulting from TNG. It also appears that in both datasets, SIMBA and TNG, the uncertainty on $\sigma_8$ is more strongly correlated with the prediction error, as evidenced by the smaller value of UCE. In general, the posterior uncertainty produced by the probabilistic model is not prone to severe miscalibration that requires mitigation, which is promising.

In SIMBA$\rightarrow$TNG scenario, whereby the encoder pre-trained on SIMBA dataset extracts the TNG dataset latent codes which are used to train/test the density estimator, the epistemic uncertainties get larger, owing to the effect of compressing out-of-distribution dataset, which then translates to the \verba{APT} ignorance. $\Omega_{\rm m}$ prediction worsens, $R^2$ = 0.927, 0.853 corresponding to SIMBA$\rightarrow$SIMBA and SIMBA$\rightarrow$TNG cases respectively, whereas $\sigma_8$ prediction in SIMBA$\rightarrow$TNG is consistent with that in SIMBA$\rightarrow$SIMBA case. Although the TNG dataset which is compressed by the $E^{\rm SIMBA}$ is out-of-distribution, the information on matter clustering is still successfully extracted by $E^{\rm SIMBA}$ such that the resulting $R^2$ obtained on $\sigma_8$ from \verba{APT} is 0.913, comparable to the one achieved in SIMBA$\rightarrow$SIMBA. In TNG$\rightarrow$SIMBA, the correlation between the ground truth and prediction gets weaker overall, although the estimator achieves $R^2 \ge 0.84$ on all cosmological parameters. Voxel value distribution of SIMBA dataset is slightly more complicated than that of TNG. As a result, the encoder built using the former tends to generalize better within the context of compressing out-of-distribution samples. This potentially explains the difference in performance of the \verba{APT} method in both scenarios, SIMBA$\rightarrow$TNG and TNG$\rightarrow$SIMBA. 

We have compared the predictive power of the posterior estimator with that of other methods, and found that in general it outperforms them all. Linear test, consisting of training a single fully connected layer on the representations without activation function, is a way to investigate how meaningful the learned representations are. The fact that the single layer network is able to predict cosmology with a precision of $R^2 \ge 0.80$ in all scenarios considered in this work points to the fact that the information carried by the featurized data is very useful. Therefore, the very promising performance of the \verba{APT} method is due to both its generalization capability and the robustness of the features retrieved by the encoders. 

Additionally, the assessment conducted on the performance of the neural density estimator, when trained/tested on latent codes of out-of-distribution dataset from a different redshift, shows that the constraining power of the method is not largely impacted within a redshift bin of 0.5. In other words, the resulting $R^2$ obtained on cosmology from the representations of a dataset at $z = 0.5$ is comparable to that of achieved on a dataset at $z = 1$ which has been used to build the encoder. This is irrespective of the simulation dataset considered. Inferring the cosmological parameters from the extracted features of a dataset at $z = 0$, on the other hand, proves to be challenging in SIMBA, especially on $\sigma_8$ which is predicted with only a precision of $R^2$ = 0.6, but the probabilistic model achieves $R^2 \sim 0.82$ on $\Omega_{\rm m}$ in the same setup. Using TNG, the cosmology is reasonably constrained by the posterior estimator at $z = 0$, as indicated by $R^2 \ge 0.80$ on all parameters.

The robustness of the encoders, which are capable of featurizing out-of-distribution samples, and the constraining power of our posterior estimator have been highlighted. The tools presented in this work can be potentially employed to analyze HI 3D tomographic data from upcoming surveys. For future work, the simulated dataset can be contaminated by noise to account for systematics that are specific to an experiment, similar to what was described in \cite{hassan2020constraining}, and different types of foregrounds. In that more realistic scenario, one can explore the robustness of the tools in terms of parameter inference. It is anticipated that dealing with noisy data is not a trivial task, regardless of the robustness of the methods presented in this work. One way to mitigate that is to resort to a domain adaptation technique. After having been trained on the simulated data (clean), the encoder is adapted to the corrupted data via some approaches like Adversarial
Discriminative Domain Adaptation \citep[ADDA; ][]{tzeng2017adversarial} or Optimal Transport \citep{courty2016optimal}.

\section*{Acknowledgments}
SA acknowledges financial support from the {\it South African Radio Astronomy Observatory} (SARAO).

%%%%%%%%%%%%%%%%%%%%%%%%%%%%%%%%%%%%%%%%%%%%%%%%%%



\bibliographystyle{mnras}
\bibliography{mnras} % if your bibtex file is called example.bib

\begin{thebibliography}{}
\makeatletter
\relax
\def\mn@urlcharsother{\let\do\@makeother \do\$\do\&\do\#\do\^\do\_\do\%\do\~}
\def\mn@doi{\begingroup\mn@urlcharsother \@ifnextchar [ {\mn@doi@}
  {\mn@doi@[]}}
\def\mn@doi@[#1]#2{\def\@tempa{#1}\ifx\@tempa\@empty \href
  {http://dx.doi.org/#2} {doi:#2}\else \href {http://dx.doi.org/#2} {#1}\fi
  \endgroup}
\def\mn@eprint#1#2{\mn@eprint@#1:#2::\@nil}
\def\mn@eprint@arXiv#1{\href {http://arxiv.org/abs/#1} {{\tt arXiv:#1}}}
\def\mn@eprint@dblp#1{\href {http://dblp.uni-trier.de/rec/bibtex/#1.xml}
  {dblp:#1}}
\def\mn@eprint@#1:#2:#3:#4\@nil{\def\@tempa {#1}\def\@tempb {#2}\def\@tempc
  {#3}\ifx \@tempc \@empty \let \@tempc \@tempb \let \@tempb \@tempa \fi \ifx
  \@tempb \@empty \def\@tempb {arXiv}\fi \@ifundefined
  {mn@eprint@\@tempb}{\@tempb:\@tempc}{\expandafter \expandafter \csname
  mn@eprint@\@tempb\endcsname \expandafter{\@tempc}}}

\bibitem[\protect\citeauthoryear{Ade et~al.,}{Ade et~al.}{2014}]{ade2014planck}
Ade P.~A.,  et~al., 2014, Astronomy \& Astrophysics, 571, A1

\bibitem[\protect\citeauthoryear{Aghanim et~al.,}{Aghanim
  et~al.}{2020}]{aghanim2020planck}
Aghanim N.,  et~al., 2020, Astronomy \& Astrophysics, 641, A1

\bibitem[\protect\citeauthoryear{Akeret, Refregier, Amara, Seehars  \&
  Hasner}{Akeret et~al.}{2015}]{akeret2015approximate}
Akeret J.,  Refregier A.,  Amara A.,  Seehars S.,   Hasner C.,  2015, Journal
  of Cosmology and Astroparticle Physics, 2015, 043

\bibitem[\protect\citeauthoryear{Alsing, Charnock, Feeney  \& Wandelt}{Alsing
  et~al.}{2019}]{alsing2019fast}
Alsing J.,  Charnock T.,  Feeney S.,   Wandelt B.,  2019, Monthly Notices of
  the Royal Astronomical Society, 488, 4440

\bibitem[\protect\citeauthoryear{Amiri et~al.,}{Amiri
  et~al.}{2022}]{amiri2022overview}
Amiri M.,  et~al., 2022, The Astrophysical Journal Supplement Series, 261, 29

\bibitem[\protect\citeauthoryear{Amiri et~al.,}{Amiri
  et~al.}{2023}]{amiri2023detection}
Amiri M.,  et~al., 2023, The Astrophysical Journal, 947, 16

\bibitem[\protect\citeauthoryear{Andrianomena \& Hassan}{Andrianomena \&
  Hassan}{2023a}]{andrianomena2023latent}
Andrianomena S.,  Hassan S.,  2023a, arXiv:2311.00799

\bibitem[\protect\citeauthoryear{Andrianomena \& Hassan}{Andrianomena \&
  Hassan}{2023b}]{andrianomena2023predictive}
Andrianomena S.,  Hassan S.,  2023b, Journal of Cosmology and Astroparticle
  Physics, 2023, 051

\bibitem[\protect\citeauthoryear{Andrianomena \& Hassan}{Andrianomena \&
  Hassan}{2025}]{andrianomena2025towards}
Andrianomena S.,  Hassan S.,  2025, Astrophysics and Space Science, 370, 14

\bibitem[\protect\citeauthoryear{Andrianomena, Hassan  \&
  Villaescusa-Navarro}{Andrianomena
  et~al.}{2025}]{andrianomena2025cosmological}
Andrianomena S.,  Hassan S.,   Villaescusa-Navarro F.,  2025, Astrophysics and
  Space Science, 370, 1

\bibitem[\protect\citeauthoryear{Ansari et~al.,}{Ansari
  et~al.}{2012}]{ansari201221}
Ansari R.,  et~al., 2012, Astronomy \& Astrophysics, 540, A129

\bibitem[\protect\citeauthoryear{Bandura et~al.,}{Bandura
  et~al.}{2014}]{bandura2014canadian}
Bandura K.,  et~al., 2014, in Ground-based and Airborne Telescopes V. pp
  738--757

\bibitem[\protect\citeauthoryear{Battye, Davies  \& Weller}{Battye
  et~al.}{2004}]{battye2004neutral}
Battye R.~A.,  Davies R.~D.,   Weller J.,  2004, Monthly Notices of the Royal
  Astronomical Society, 355, 1339

\bibitem[\protect\citeauthoryear{Battye, Browne, Dickinson, Heron, Maffei  \&
  Pourtsidou}{Battye et~al.}{2013}]{battye2013h}
Battye R.,  Browne I.,  Dickinson C.,  Heron G.,  Maffei B.,   Pourtsidou A.,
  2013, Monthly Notices of the Royal Astronomical Society, 434, 1239

\bibitem[\protect\citeauthoryear{Bharadwaj \& Pandey}{Bharadwaj \&
  Pandey}{2005}]{bharadwaj2005probing}
Bharadwaj S.,  Pandey S.~K.,  2005, Monthly Notices of the Royal Astronomical
  Society, 358, 968

\bibitem[\protect\citeauthoryear{Binnie, Zhao, Pritchard  \& Mao}{Binnie
  et~al.}{2025}]{binnie2025likelihood}
Binnie T.,  Zhao X.,  Pritchard J.,   Mao Y.,  2025, arXiv:2502.08152

\bibitem[\protect\citeauthoryear{Bishop}{Bishop}{1994}]{bishop1994mixture}
Bishop C.~M.,  1994, Aston University

\bibitem[\protect\citeauthoryear{Blum}{Blum}{2010}]{blum2010approximate}
Blum M.~G.,  2010, Journal of the American Statistical Association, 105, 1178

\bibitem[\protect\citeauthoryear{Bull, Ferreira, Patel  \& Santos}{Bull
  et~al.}{2015}]{bull2015late}
Bull P.,  Ferreira P.~G.,  Patel P.,   Santos M.~G.,  2015, The Astrophysical
  Journal, 803, 21

\bibitem[\protect\citeauthoryear{Child}{Child}{2020}]{child2020very}
Child R.,  2020, arXiv preprint arXiv:2011.10650

\bibitem[\protect\citeauthoryear{Courty, Flamary, Tuia  \&
  Rakotomamonjy}{Courty et~al.}{2016}]{courty2016optimal}
Courty N.,  Flamary R.,  Tuia D.,   Rakotomamonjy A.,  2016, IEEE transactions
  on pattern analysis and machine intelligence, 39, 1853

\bibitem[\protect\citeauthoryear{Crichton et~al.,}{Crichton
  et~al.}{2022}]{crichton2022hydrogen}
Crichton D.,  et~al., 2022, Journal of Astronomical Telescopes, Instruments,
  and Systems, 8, 011019

\bibitem[\protect\citeauthoryear{Dav{\'e}, Angl{\'e}s-Alc{\'a}zar, Narayanan,
  Li, Rafieferantsoa  \& Appleby}{Dav{\'e} et~al.}{2019}]{dave2019simba}
Dav{\'e} R.,  Angl{\'e}s-Alc{\'a}zar D.,  Narayanan D.,  Li Q.,  Rafieferantsoa
  M.~H.,   Appleby S.,  2019, Monthly Notices of the Royal Astronomical
  Society, 486, 2827

\bibitem[\protect\citeauthoryear{Dawson et~al.,}{Dawson
  et~al.}{2016}]{dawson2016sdss}
Dawson K.~S.,  et~al., 2016, The Astronomical Journal, 151, 44

\bibitem[\protect\citeauthoryear{Eickenberg, Exarchakis, Hirn  \&
  Mallat}{Eickenberg et~al.}{2017}]{eickenberg2017solid}
Eickenberg M.,  Exarchakis G.,  Hirn M.,   Mallat S.,  2017, Advances in Neural
  Information Processing Systems, 30

\bibitem[\protect\citeauthoryear{Eickenberg, Exarchakis, Hirn, Mallat  \&
  Thiry}{Eickenberg et~al.}{2018}]{eickenberg2018solid}
Eickenberg M.,  Exarchakis G.,  Hirn M.,  Mallat S.,   Thiry L.,  2018, The
  Journal of chemical physics, 148

\bibitem[\protect\citeauthoryear{Gal \& Ghahramani}{Gal \&
  Ghahramani}{2016}]{gal2016dropout}
Gal Y.,  Ghahramani Z.,  2016, in international conference on machine learning.
  pp 1050--1059

\bibitem[\protect\citeauthoryear{Germain, Gregor, Murray  \&
  Larochelle}{Germain et~al.}{2015}]{germain2015made}
Germain M.,  Gregor K.,  Murray I.,   Larochelle H.,  2015, in International
  conference on machine learning. pp 881--889

\bibitem[\protect\citeauthoryear{Gillet, Mesinger, Greig, Liu  \& Ucci}{Gillet
  et~al.}{2019}]{gillet2019deep}
Gillet N.,  Mesinger A.,  Greig B.,  Liu A.,   Ucci G.,  2019, Monthly Notices
  of the Royal Astronomical Society, 484, 282

\bibitem[\protect\citeauthoryear{Greenberg, Nonnenmacher  \& Macke}{Greenberg
  et~al.}{2019}]{greenberg2019automatic}
Greenberg D.,  Nonnenmacher M.,   Macke J.,  2019, in International conference
  on machine learning. pp 2404--2414

\bibitem[\protect\citeauthoryear{Greig \& Mesinger}{Greig \&
  Mesinger}{2015}]{greig201521cmmc}
Greig B.,  Mesinger A.,  2015, Monthly Notices of the Royal Astronomical
  Society, 449, 4246

\bibitem[\protect\citeauthoryear{Greig, Ting  \& Kaurov}{Greig
  et~al.}{2023}]{greig2023detecting}
Greig B.,  Ting Y.-S.,   Kaurov A.~A.,  2023, Monthly Notices of the Royal
  Astronomical Society, 519, 5288

\bibitem[\protect\citeauthoryear{Hassan, Andrianomena  \& Doughty}{Hassan
  et~al.}{2020}]{hassan2020constraining}
Hassan S.,  Andrianomena S.,   Doughty C.,  2020, Monthly Notices of the Royal
  Astronomical Society, 494, 5761

\bibitem[\protect\citeauthoryear{Ho}{Ho}{1995}]{ho1995random}
Ho T.~K.,  1995, in Proceedings of 3rd international conference on document
  analysis and recognition. pp 278--282

\bibitem[\protect\citeauthoryear{Jennings \& Madigan}{Jennings \&
  Madigan}{2017}]{jennings2017astroabc}
Jennings E.,  Madigan M.,  2017, Astronomy and computing, 19, 16

\bibitem[\protect\citeauthoryear{Jonas}{Jonas}{2009}]{jonas2009meerkat}
Jonas J.~L.,  2009, Proceedings of the IEEE, 97, 1522

\bibitem[\protect\citeauthoryear{Laves, Ihler, Fast, Kahrs  \& Ortmaier}{Laves
  et~al.}{2020}]{laves2020well}
Laves M.-H.,  Ihler S.,  Fast J.~F.,  Kahrs L.~A.,   Ortmaier T.,  2020, in
  Medical imaging with deep learning. pp 393--412

\bibitem[\protect\citeauthoryear{Lueckmann, Goncalves, Bassetto, {\"O}cal,
  Nonnenmacher  \& Macke}{Lueckmann et~al.}{2017}]{lueckmann2017flexible}
Lueckmann J.-M.,  Goncalves P.~J.,  Bassetto G.,  {\"O}cal K.,  Nonnenmacher
  M.,   Macke J.~H.,  2017, Advances in neural information processing systems,
  30

\bibitem[\protect\citeauthoryear{Majumdar, Pritchard, Mondal, Watkinson,
  Bharadwaj  \& Mellema}{Majumdar et~al.}{2018}]{majumdar2018quantifying}
Majumdar S.,  Pritchard J.~R.,  Mondal R.,  Watkinson C.~A.,  Bharadwaj S.,
  Mellema G.,  2018, Monthly Notices of the Royal Astronomical Society, 476,
  4007

\bibitem[\protect\citeauthoryear{Martini et~al.,}{Martini
  et~al.}{2018}]{martini2018overview}
Martini P.,  et~al., 2018, in Ground-based and Airborne Instrumentation for
  Astronomy VII. pp 410--420

\bibitem[\protect\citeauthoryear{McInnes, Healy  \& Melville}{McInnes
  et~al.}{2018}]{mcinnes2018umap}
McInnes L.,  Healy J.,   Melville J.,  2018, arXiv:1802.03426

\bibitem[\protect\citeauthoryear{Mellema et~al.,}{Mellema
  et~al.}{2013}]{mellema2013reionization}
Mellema G.,  et~al., 2013, Experimental Astronomy, 36, 235

\bibitem[\protect\citeauthoryear{Mellema, Koopmans, Shukla, Datta, Mesinger  \&
  Majumdar}{Mellema et~al.}{2015}]{mellema2015hi}
Mellema G.,  Koopmans L.,  Shukla H.,  Datta K.~K.,  Mesinger A.,   Majumdar
  S.,  2015, arXiv preprint arXiv:1501.04203

\bibitem[\protect\citeauthoryear{Nelson et~al.,}{Nelson
  et~al.}{2019}]{nelson2019illustristng}
Nelson D.,  et~al., 2019, Computational Astrophysics and Cosmology, 6, 1

\bibitem[\protect\citeauthoryear{Novaes et~al.,}{Novaes
  et~al.}{2024}]{novaes2024cosmological}
Novaes C.~P.,  et~al., 2024, Monthly Notices of the Royal Astronomical Society,
  528, 2078

\bibitem[\protect\citeauthoryear{Papamakarios \& Murray}{Papamakarios \&
  Murray}{2016}]{papamakarios2016fast}
Papamakarios G.,  Murray I.,  2016, Advances in neural information processing
  systems, 29

\bibitem[\protect\citeauthoryear{Papamakarios, Pavlakou  \&
  Murray}{Papamakarios et~al.}{2017}]{papamakarios2017masked}
Papamakarios G.,  Pavlakou T.,   Murray I.,  2017, Advances in neural
  information processing systems, 30

\bibitem[\protect\citeauthoryear{Papamakarios, Sterratt  \&
  Murray}{Papamakarios et~al.}{2019}]{papamakarios2019sequential}
Papamakarios G.,  Sterratt D.,   Murray I.,  2019, in The 22nd international
  conference on artificial intelligence and statistics. pp 837--848

\bibitem[\protect\citeauthoryear{Pourtsidou, Bacon  \& Crittenden}{Pourtsidou
  et~al.}{2017}]{pourtsidou2017h}
Pourtsidou A.,  Bacon D.,   Crittenden R.,  2017, Monthly Notices of the Royal
  Astronomical Society, 470, 4251

\bibitem[\protect\citeauthoryear{Rasmussen \& Williams}{Rasmussen \&
  Williams}{2006}]{williams2006gaussian}
Rasmussen C.,  Williams C.,  2006, Gaussian Processes for Machine Learning.
Adaptive Computation and Machine Learning, MIT Press, Cambridge, MA, USA

\bibitem[\protect\citeauthoryear{Rezende \& Mohamed}{Rezende \&
  Mohamed}{2015}]{rezende2015variational}
Rezende D.,  Mohamed S.,  2015, in International conference on machine
  learning. pp 1530--1538

\bibitem[\protect\citeauthoryear{Smoot \& Debono}{Smoot \&
  Debono}{2017}]{smoot201721}
Smoot G.~F.,  Debono I.,  2017, Astronomy \& Astrophysics, 597, A136

\bibitem[\protect\citeauthoryear{Switzer et~al.,}{Switzer
  et~al.}{2013}]{switzer2013determination}
Switzer E.,  et~al., 2013, Monthly Notices of the Royal Astronomical Society:
  Letters, 434, L46

\bibitem[\protect\citeauthoryear{Tejero-Cantero, Boelts, Deistler, Lueckmann,
  Durkan, Gon{\c{c}}alves, Greenberg  \& Macke}{Tejero-Cantero
  et~al.}{2020a}]{tejero2020sbi}
Tejero-Cantero A.,  Boelts J.,  Deistler M.,  Lueckmann J.-M.,  Durkan C.,
  Gon{\c{c}}alves P.~J.,  Greenberg D.~S.,   Macke J.~H.,  2020a,
  arXiv:2007.09114

\bibitem[\protect\citeauthoryear{Tejero-Cantero, Boelts, Deistler, Lueckmann,
  Durkan, Gonçalves, Greenberg  \& Macke}{Tejero-Cantero
  et~al.}{2020b}]{tejero-cantero2020sbi}
Tejero-Cantero A.,  Boelts J.,  Deistler M.,  Lueckmann J.-M.,  Durkan C.,
  Gonçalves P.~J.,  Greenberg D.~S.,   Macke J.~H.,  2020b, \mn@doi [Journal
  of Open Source Software] {10.21105/joss.02505}, 5, 2505

\bibitem[\protect\citeauthoryear{Tzeng, Hoffman, Saenko  \& Darrell}{Tzeng
  et~al.}{2017}]{tzeng2017adversarial}
Tzeng E.,  Hoffman J.,  Saenko K.,   Darrell T.,  2017, in Proceedings of the
  IEEE conference on computer vision and pattern recognition. pp 7167--7176

\bibitem[\protect\citeauthoryear{Villaescusa-Navarro
  et~al.,}{Villaescusa-Navarro et~al.}{2021a}]{villaescusa2021multifield}
Villaescusa-Navarro F.,  et~al., 2021a, arXiv:2109.09747

\bibitem[\protect\citeauthoryear{Villaescusa-Navarro
  et~al.,}{Villaescusa-Navarro et~al.}{2021b}]{villaescusa2021camels}
Villaescusa-Navarro F.,  et~al., 2021b, The Astrophysical Journal, 915, 71

\bibitem[\protect\citeauthoryear{Villaescusa-Navarro
  et~al.,}{Villaescusa-Navarro et~al.}{2022a}]{pacodatarelease}
Villaescusa-Navarro F.,  et~al., 2022a, arXiv:2201.01300

\bibitem[\protect\citeauthoryear{Villaescusa-Navarro
  et~al.,}{Villaescusa-Navarro et~al.}{2022b}]{villaescusa2022camels}
Villaescusa-Navarro F.,  et~al., 2022b, The Astrophysical Journal Supplement
  Series, 259, 61

\bibitem[\protect\citeauthoryear{Wagg, Bourke, Green, Braun  et~al.}{Wagg
  et~al.}{2014}]{wagg2014ska1}
Wagg J.,  Bourke T.,  Green J.,  Braun R.,   et~al., 2014, SKA1 scientific use
  cases

\bibitem[\protect\citeauthoryear{Zhao, Mao, Cheng  \& Wandelt}{Zhao
  et~al.}{2022}]{zhao2022simulation}
Zhao X.,  Mao Y.,  Cheng C.,   Wandelt B.~D.,  2022, The Astrophysical Journal,
  926, 151

\bibitem[\protect\citeauthoryear{Zhao, Mao, Zuo  \& Wandelt}{Zhao
  et~al.}{2024}]{zhao2024simulation}
Zhao X.,  Mao Y.,  Zuo S.,   Wandelt B.~D.,  2024, The Astrophysical Journal,
  973, 41

\makeatother
\end{thebibliography}

%\begin{thebibliography}{99}
%\end{thebibliography}

%%%%%%%%%%%%%%%%%%%%%%%%%%%%%%%%%%%%%%%%%%%%%%%%%%

%%%%%%%%%%%%%%%%% APPENDICES %%%%%%%%%%%%%%%%%%%%%
\onecolumn
\appendix

%%%%%%%%%%%%%%%%%%%%%%%%%%%%%%%%%%%%%%%%%%%%%%%%%%

% Don't change these lines
\bsp	% typesetting comment
\label{lastpage}
\end{document}